\documentclass[seceq]{ptptex}

\usepackage{graphicx}

\title{
The Pion Cloud: Insights into Hadron Structure %
}

\author{
Anthony W. \textsc{Thomas}
}

\inst{
Jefferson Lab, 12000 Jefferson Ave., Newport News VA 23606 USA  and \\
College of William and Mary, Williamsburg VA 23187 USA
}

\abst{
Modern nuclear theory presents a fascinating study in contrasting
approaches to the structure of hadrons and nuclei. Nowhere is this
more apparent than in the treatment of the pion cloud. As this
discussion really begins with Yukawa,
it is entirely appropriate that this invited
lecture at the Yukawa Institute in Kyoto should deal with the issue.
}

\begin{document}

\maketitle

\section{Introduction}
While the pion has been with us in theory since 1935 and in fact 
since 1947, chiral symmetry has 
provided important guidance in dealing with strong interaction 
phenomena since the 1960s. Well known work by Gell Mann and 
L\'evy, Weinberg, Pagels and many others developed the consequences before 
the underlying origins in terms of QCD were discovered -- 
for a review see Ref.~\cite{Pagels:1974se}. That early 
work was not only mathematically rigorous but was guided by sound 
physics insight. Since the early 1980s a great deal of work which 
exploits the chiral symmetry of QCD has been carried out within 
the framework of chiral perturbation theory. This work is usually 
descibed in language which equates it to QCD and talks about a 
systematic, rigorous description of the hadronic world in terms 
of a set of low energy constants.
In a few outstanding examples this approach has proven very 
valuable indeed. For some low energy phenomena it has proven possible 
to make rigorous statements concerning cross sections that had previously 
seemed incurably model dependent. 

Yet, the rise of chiral perturbation 
theory ($\chi$PT) has something of a dark side too.
At its best, within a well defined 
radius of convergence in an expansion in powers of pion mass ($m_\pi$) 
or momentum transfer ($Q$), $\chi$PT provides a mathematically rigorous 
connection between data sets. Low energy constants derived by fitting one 
set of data lead to unambiguous predictions elsewhere. In the reductio ad 
absurdum limit, the low energy world would be 
reduced to engineering. One would
look up a set of low energy constants in the standard reference and turn 
the handle of a well defined formalism to calculate the result of a new 
measurement. The low energy constants are by and large without physical
content, in many cases scheme or renormalization scale dependent. No 
physical intuition is either possible or necessary in order to use the 
formalism. Many physicists find such a world at best dreary or at 
worst deathly dull. Having as a major aim in life the determination of 
the as yet unknown (say) 97th low energy constant in a never ending series 
of physically meaningless parameters does not make getting out of bed in 
the morning a great joy -- for most physicists.

On a more practical level, the systematic order by order expansion 
involving rigorous application of a strict counting scheme 
is only meaningful if one lies within the the corresponding region 
of convergence. Yet 
the rigor of going to higher order in a divergent series is rarely discussed. 
The answer at N$^3$LO is not necessarily better than NLO in that case, 
although it may take several Ph.~D.\ students to get there. The rigor of 
working consistently within a well defined counting scheme is ephemeral if 
the counting scheme itself is based on inadequate physics -- 
involving, for example, 
the wrong degrees of freedom. Classic examples of the latter include studies 
of spin dependent quantities which define a counting scheme that rigorously 
excludes the $\Delta$ resonance.    

Of course, $\chi$PT itself was in part a natural reaction to 
the lawlessness of some of the model building that preceded it, where 
theorists acting more like snake oil salesmen argued the merits of their 
particular brand in the absence of objective criteria by which to judge.
As a particularly extreme reaction to this, some practitioners of $\chi$PT 
go so far as to classify the concept of a pion cloud 
surrounding hadrons as so ill-defined as to be vigorously eschewed.

In the context of this stimulating meeting at Yukawa's institute it seems 
worthwhile to re-examine the concept of the pion cloud. 
In doing so we will find that it has 
indeed proven its value. Time and again it has predicted new results, 
later confirmed by experiment or more recently by numerical experiments 
with QCD itself. It has also provided simple and 
physically clear and meaningful 
interpretations of quite diverse phenomena 
and is still making new predictions 
which will only be addressed by the next generation of experimental facilities.

\section{A Framework for Understanding the Role of the Pion Cloud in Hadron 
Structure}
Amongst the many models developed to incorporate the constraints of chiral 
symmetry into a quantitative picture of hadron structure the cloudy bag model 
(CBM) was unique~\cite{Theberge:1980ye,Thomas:1981vc,Thomas:1982kv}. 
It did not attempt to describe the short distance structure 
in terms of pions, since the natural degrees of freedom in QCD are the quarks 
and gluons. It simply introduced the pion field on top of a successful model of 
quark confinement (the MIT bag model), in the simplest way consistent with 
chiral symmetry. It was possible to prove rigorously that, provided the size 
of the bag (read confinement region) was sufficiently 
large (say 0.8fm or more), 
the pion cloud could be treated perturbatively~\cite{Dodd:1981ve} 
-- the size of the quark core 
provided a natural mechanism for suppressing the emission of more than one 
or two pions at a time.

Although the CBM began with the minimal coupling of the pion field to the 
surface of the bag confining the valence quarks necessary to restore 
chiral symmetry, it was soon transformed 
into a form with pseudo-vector coupling 
throughout the bag volume~\cite{Thomas:1981ps}, which 
made it much easier to derive some crucial low energy theorems, such as the 
Weinberg-Tomozawa formula. (It was also discovered that, at least for massless 
$u$ and $d$ quarks, the results derived using surface coupling were valid in 
the volume coupled version. For example, the single pion exchange force between 
two nucleons was identical to that between point-like nucleons for separations 
greater than twice the bag radius ($r>2R$), with the form factor 
arising from finite 
nucleon size only modifying it when the bags overlapped, in {\it both} 
formulations.) The chiral quark model of 
Manohar and Georgi~\cite{Manohar:1983md}, 
which was published a few years later, was very similar to that 
version except that it did not carry out the projection onto 
colorless, bare hadronic states (the P-space 
projection)~\cite{Thomas:1982kv}. Only a decade later 
was it realized~\cite{Jenkins:1991es} that this projection 
is essential to ensure the correct infrared limit and hence the correct 
leading and next-to-leading non-analytic (LNA and NLNA) behaviour of 
various hadronic properties~\cite{Thomas:1999mu}. 
Unfortunately, this lesson is still not widely 
appreciated and many calculations 
within so-called chiral quark models are still carried out in a manner which 
actually violates chiral symmetry! 

We now review a large number of important physical 
results which follow directly 
from the perturbative nature of the pion cloud found naturally within 
the cloudy bag model.

\subsection{Neutron Charge Distribution}
Within the CBM one had at once a very natural and 
beautiful way to understand the charge 
distribution of the neutron~\cite{Thomas:1981vc}. 
It had of course been known that the long range 
negative tail of the neutron was explained in terms of pion emission. But 
within the Chew-Low model the expansion in terms of the number of pions was 
divergent and one could say nothing about the short distance structure 
(inside 1~fm). 
In the CBM it was suddenly simple; the neutron could be thought of as 
predominantly a bare neutron bag (with zero charge distribution), but 
occasionally as a $\pi^-$ and a bare proton bag whose charge distribution 
was trivial to calculate. Furthermore, the structure of the pion cloud was 
such that it peaked in the bag surface, so that the peak of the negative 
charge distribution of the neutron had a natural interpretation in 
terms of the bag surface (or the size of the volume within which the valence 
quarks were confined). On a personal note, I first realized this on New Year's
Eve 1980 and had to wait in considerable anticipation for many years 
before experimental measurements were able 
to pin down the position of this peak. 
Modern recoil polarization measurements have established 
that this peak is indeed around 0.8~fm~\cite{Kelly:2003zu}, 
as anticipated in the CBM.

The studies of many different phenomena within the CBM are consistent with 
a picture in which the total probability of a physical nucleon consisting 
of a single pion and a bare nucleon is 
approximately 20\% (and a pion and a bare
$\Delta$ around 8-10\%)~\cite{Speth:1996pz}. For the proton 
(rounding 20\% to 21\% to simplify the algebra) 
this means roughly 14\% $\pi^+$-bare n and 
7\% $\pi^0$-bare p. Similarly for the neutron it's 14\% $\pi^-$-bare p and 7\% 
$\pi^0$-bare n. Taking the charge density at the 
centre of a bare p to be $x$ and 
at the center of a bare n to be 0, 
this gives for the ratio of the central charge 
densities of the physical n to p to be approximately $0.14 x /(0.79 x + 0.07 x)
= 0.14/0.86 \sim 1/6$, which is very close to the experimental ratio extracted 
from modern studies of the neutron electric form factor at 
Nikhef, Mainz, 
MIT-Bates 
and JLab~\cite{Passchier:1999cj,Bermuth:2003qh,Zhu:2001md,Kelly:2003zu}. Small 
corrections may be expected from the 
$\pi \Delta$ component of the wave function 
and from hyperfine effects, which may result in a slightly non-zero charge 
distribution in the bare n, but the result just derived in this simple 
and physically transparent manner remains essentially correct.
    
\subsection{``Flavor Symmetry'' Violations in the Nucleon Sea}
Feynman's much appreciated physical insight led to calculations in the early 
1970s of a possible pion contribution to the 
nucleon sea~\cite{Sullivan:1971kd}. However, this was 
largely ignored until the discovery of the famous EMC effect generated 
enormous interest in the possible role of an 
excess of pions associated with nuclear 
binding on the structure functions of 
nuclei~\cite{Ericson:1983um,LlewellynSmith:1983qa}. 
At this time it was realized that 
the presence of a pion cloud around 
the nucleon would have profound consequences 
for the flavor structure of the proton sea. In particular, with the biggest 
pionic component of the proton wave function being $\pi^+$n and the $\pi^+$ 
containing only down anti-quarks, it was clear that one expected an excess 
of anti-down over anti-up quarks in the nucleon sea~\cite{Thomas:1983fh}. 
Using the probabilites 
noted earlier one expected an excess of about 0.14 anti-down quarks. 

At the time this novel property of the nucleon sea was predicted 
it was extremely unusual to think about deep 
inelastic scattering in terms of hadronic Fock components and certainly not 
in terms of a pion cloud. It was not until the experimental 
confirmation of a violation of the 
Gottfried sum-rule almost 8 years later~\cite{Amaudruz:1991at} 
that this began to change.  Early 
discussions of the violation of the Gottfried sum-rule often talked about 
flavor symmetry violation and appeared to confuse that with a violation of 
isospin, or charge independence. In fact, the simple explanation of how this 
excess of anti-down quarks was 
predicted~\cite{Thomas:1983fh,Signal:1991ug} makes 
it clear that it has {\it nothing} 
to do with isospin violation. The 2:1 ratio of the $\pi^+$n to $\pi^0$p Fock 
components is a direct consequence of isospin symmetry. In contrast, the term  
``flavor symmetry'' was at best vague and at worst misleading. 

In summary, the concept of a pion cloud surrounding the proton led to a simple 
and natural prediction of an excess of anti-down quarks in the proton sea. 
The prediction was not only qualitatively but quantitatively in agreement with 
experiments performed many years later. This discovery has completely changed 
the standard fits to parton distribution functions.   

Perhaps surprisingly, the original paper which predicted the excess of 
anti-down sea quarks was not primarily concerned with that problem. Its main 
subject was actually to point out the considerable difference between the 
strength of the strange and non-strange sea 
predicted by the meson cloud picture. 
Indeed, with the hadron size providing a natural high momentum cut-off on the 
meson-baryon dynamics, it was observed that the strange sea arising from the 
fluctuation p $\rightarrow K^+ \Lambda$ would necessarily be much smaller than 
the non-strange sea associated with the pion cloud~\cite{Thomas:1983fh}. 
This was proposed as a 
natural explanation of the 2:1 ratio that had already been seen in neutrino 
deep inelastic scattering~\cite{Abramowicz:1982zr}. 
Of course, as well as this sea 
arising from the meson clouds, which 
is non-perturbative in the usual sense in which one discusses QCD, there is 
necessarily a perturbative sea generated through the process $g \rightarrow q + 
\bar{q}$ which would be approximately flavor blind 
(at least for $u, \, d$ and 
$s$).

The consequences for the strange sea of the nucleon were explored further by 
Signal and Thomas just a few years later~\cite{Signal:1987gz}. 
It was noted that while the number of 
$s$ and $\bar{s}$ quarks must be equal in the proton, 
their momentum distributions 
will in general not be the same~\cite{Melnitchouk:1999mv}. 
This is essentially because of the unequal sharing 
of light-cone momentum between the $K^+$ and $\Lambda$ in the corresponding 
piece of the proton wave function. 
Although we still have no experimental guidance on this 
issue, it has turned out to be very important in the context of using 
neutrino deep inelastic scattering to test the Standard Model. 
It is clearly very important to reliably and accurately determine 
at least the integral 
of $x ( s(x) - \bar{s}(x) )$ as soon as possible. 
    
\subsection{Role of the Decuplet in Octet Magnetic Moments}
The first systematic inclusion of the decuplet baryons in a calculation of 
octet magnetic moments was in the study by 
Th\'eberge and Thomas~\cite{Theberge:1982xs}. On 
physical grounds it was clear that any spin flip quantity would be sensitive 
to the inclusion of the decuplet. The very convergence of the expansion in 
pion number for the nucleon was directly related to the role of the $\Delta$ 
in the vertex renormalization of the 
pion-nucleon coupling constant~\cite{Thomas:1982kv,Dodd:1981ve}. The 
results for the octet magnetic moments were in very reasonable agreement 
with experiment. Within the context of formal $\chi$PT it was a decade 
before decuplet contributions were incorporated in the analysis and even then 
it took a while to realize that the correct LNA structure required a projection 
onto bare-baryon pion configurations. 

\subsection{Physically Transparent Estimate of the Strange Quark 
Content of the Proton}
We note first that there is no known example where 
the current quark masses show up in hadron physics undressed 
by non-perturbative glue. Thus the cost 
to make an $s-\bar{s}$ pair in the proton is of order 1.0 to 1.1 GeV 
(twice the strange constituent quark mass). 
On the other hand, creating the $\bar{s}$ in a kaon and the $s$ in a 
$\Lambda$ costs only 0.65 GeV. (Note that the N to $K \Sigma$ coupling 
is considerably smaller than that for 
N to $K \Lambda$ and hence, in this simple 
discussion, we ignore it.) On these grounds alone we expect the virtual 
transition N to $K \Lambda$ to dominate the production of strangeness 
in the proton. 

The probability for finding the $K \Lambda$ configuration is inversely 
proportional to the excitation energy squared. Naively the transition 
N to N $\pi$ costs 140 MeV but with additional kinetic energy this is 
around 600 MeV in total. Including 
kinetic energy for the $K \Lambda$ component as well, it costs 
roughly twice as much as N $\pi$. Thus the $K \Lambda$ probability is 
expected to be of order 5\%.

We consider first the strangeness radius of the proton, based on 
this 5\% $K \Lambda$ probability. In the CBM the radius of a $\Lambda$ 
bag is about 1~fm, which yields a mean square radius for the 
strange quark around 0.5~fm$^2$.  
As an estimate of the range of variation possible, we also take the bag 
radius $R = 0.8$~fm, with a corresponding mean square radius 
close to 0.32~fm$^2$. 
In order to estimate the contribution from the kaon cloud, we need to realize 
that the peak in the Goldstone boson wave function 
is at the confinement (bag) radius~\cite{Lu:1997sd,Thomas:2005qm}. 
The meson field then decreases with a 
range between one over the energy cost of the Fock state 
and $1/(m_K +m_\Lambda - m_N)$. Thus for $R=0.8$ fm we get a mean square 
radius for the $\bar{s}$ 
distribution of order 1 fm$^2$, while for $R = 1$fm we get about 1.4fm$^2$. 
Weighting the $s$ by $-1/3$ and $\bar{s}$ by $+1/3$, 
we find that the mean square 
charge radius of strange quarks is between $(-0.32 + 1.0)/3$ and 
$(-0.5 +1.4)/3$; that is, in the range (0.23,0.30) fm$^2$, 
times the probability for finding the $K \Lambda$ configuration. 

To calculate $G_E^s$ at $Q^2=0.1$ GeV$^2$ = 2.5 fm$^{-2}$, 
we assume that the term \\
$- Q^2  \langle r^2 \rangle / 6$ 
dominates and finally multiply 
by $-3$ to agree with the usual convention of 
removing the strange quark charge. 
This yields $G_E^s \in (+0.01,+0.02)$. It is definitely small and definitely 
positive for the very clear physical reasons that the $K \Lambda$ probability 
is small and that the kaon cloud extends outside the $\Lambda$. 
A comparison with the currently preferred fit~\cite{Young:2006jc} 
to the existing world 
data~\cite{Armstrong:2005hs,Aniol:2005zg,Aniol:2005zf,Maas:2004dh} 
reveals that this estimate has the opposite sign. 
It is also significantly smaller in magnitude. 
However, given the current experimental 
errors, the agreement with data is excellent. (We also note the significant 
charge symmetry correction in $^4$He reported recently, 
which would tend to move the central 
value of the experiment closer to theory~\cite{Viviani:2007gi}.) 
Finally, we note that this simple 
estimate of the mean square strange radius is also in excellent agreement with 
the recent calculation based on lattice
QCD~\cite{Leinweber:2006ug}, namely 
$G_E^s(0.1 \, {\rm GeV}^2) = 0.001 \pm 0.004 \pm 0.004 \, {\rm fm}^2$. 

Because orbital angular momentum is quantized, the contribution to the magnetic 
moment from the $\bar{s}$ in the kaon cloud is much less model dependent. The 
Clebsch-Gordon coefficients show that in a spin-up proton the 
probability of a spin down (up) $\Lambda$, accompanied by a kaon with orbital 
angular momentum +1 (0), is 2/3 (1/3). We also know the magnetic moment 
of the $\Lambda$ and that it is dominated by the magnetic moment of the 
$s$ quark. Hence the total strangeness magnetic moment of the proton is 
$-3 \times P_{K \Lambda} \times 2/3 \times (+0.6 +1/3) -3 
\times P_{K \Lambda} \times 1/3 \times (-0.6 + 0)$, 
where the terms in parentheses are, respectively, the magnetic moment of the 
spin down (up) $\Lambda$ and the magnetic moment of the charge 
+1/3 $\bar{s}$ quark with one unit (or zero units) of orbital angular momentum. 
The net result, namely $G_M^s = - 0.063 \mu_N$, is reasonably close to the 
most recent lattice QCD estimate~\cite{Leinweber:2004tc}, that is,  
$G_M^s  = -0.046 \pm 0.019 \, \mu_N$. From the point of 
view of this ``back of the envelope'' estimate, the lattice result clearly 
has both a natural magnitude and sign. 

\section{Discoveries in Modern Lattice QCD}
One of the unexpected but very positive consequences of our {\it lack} of 
supercomputing power is the fact that it has not been possible to compute 
physical hadron properties in lattice QCD. In fact, with computation time 
scaling like $m_\pi^{-9}$ (if we include the larger lattice size needed), 
calculations have covered the pion mass range from 0.3 to 1.0 GeV (or higher).
Far from being a disappointment, this has given us a wealth of unexpected 
insight into how QCD behaves as the light quark masses 
are varied~\cite{Thomas:2002sj}. In terms of 
the insight this has given us into hadron structure it is both truly invaluable 
and thus far under-utilized. 

The most striking feature of the lattice data is that in the region 
$m_\pi > 0.4$ GeV, in fact for almost all of the simulations made so far, 
all baryon properties show a smooth dependence on 
quark mass, totally consistent 
with a constituent quark model. The rapid, non-linear dependence on $m_\pi$ 
required by the LNA and NLNA behavior of $\chi$PT are notably absent from the 
data!

The conventional view of $\chi$PT has no explanation for this simple, universal 
observation. 
Worse, in seeking to apply $\chi$PT to extrapolate the data back to 
the physical pion mass, it has been necessary to rely on ad hoc cancellations 
between the high order terms in the usual power series expansion (supplemented 
by the  required non-analytic behavior). In fact, there is strong evidence that 
such series expansions have been applied well 
beyond their region of convergence~\cite{Leinweber:2005xz}
and that as a result the extrapolations are largely unreliable.
\begin{figure}[b]
\vspace{-0.5cm}
\centering
\includegraphics[width=14.0cm]{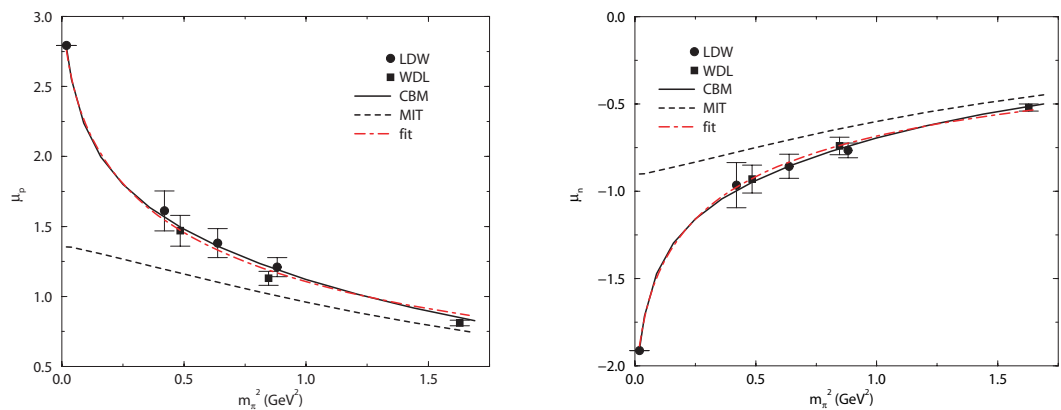}
\caption{Comparison between early lattice data for $\mu_p$ and $\mu_n$ versus 
$m_\pi^2$ and the CBM~\protect\cite{Leinweber:1998ej}.
The model is compared with a simple chiral fit and the MIT bag
model (without the pion cloud).
}\label{MagMom}
\end{figure}

On the other hand, the picture of the pion cloud 
that we have presented here yields 
an extremely natural explanation of the universal, 
constituent quark model behaviour 
of hadron properties found in the lattice simulations for $m_\pi > 0.4$ GeV. 
The natural high momentum cut-off on the momentum of the emitted pion, 
which is associated with the 
finite size (typically $R \sim$ 1 fm) of the bare baryon 
(i.e., the bag in the CBM), strongly 
suppresses pion loop contributions as $m_\pi$ increases. 
The natural mass scale which 
sets the boundary between rapid chiral variation and constituent quark type 
behavior is $1/R \sim 0.2 \, {\rm to} \, 0.4$ GeV. 
Indeed, when in the early investigation of the 
quark mass dependence of nucleon properties the CBM was 
compared directly with lattice 
data, the agreement was remarkably good~\cite{Leinweber:1999ig}. 
(Similar results have been obtained recently within the chiral quark soliton 
model~\cite{Goeke:2005fs}.)
We illustrate this in Fig.~\ref{MagMom} for 
the magnetic moments of the proton and neutron. 
The results were equally as impressive for 
the N and $\Delta$ masses and 
magnetic moments, the proton charge radius and the 
moments of its parton distribution 
functions~\cite{Detmold:2002nf}. The key features necessary to reproduce 
the behaviour found at large quark mass 
in lattice QCD {\em and} to reproduce the experimentally 
measured data at the physical mass 
seem to be that:
\begin{itemize}
\item The treatment of the pion cloud (chiral) 
corrections ensures the correct LNA 
(and NLNA, although in practice this seems less 
important in many applications) behaviour of QCD 
\item The pion cloud contribution is 
suppressed for $m_\pi$ beyond 0.4 GeV, and
\item the underlying quark model exhibits constituent 
quark like behaviour for the 
corresponding range of current quark masses.
\end{itemize}
The CBM satisfies all of these properties.    

In practice, of course, in analyzing lattice QCD data 
one does not want to rely on any 
particular quark model. However, one does need to suppress the pion cloud as 
$m_\pi$ goes up, and the simple use of a finite range regulator 
(FRR) in the evaluation of 
the pion loops that yield the LNA and NLNA behaviour ensures 
this at the cost of one 
additional parameter, the cut-off mass $\Lambda$. 
If the data are good enough one can 
use this as a fitting parameter but in general 
it is sufficient~\cite{Lepage:1997cs} to choose a value 
consistent with the physical arguments presented 
above (e.g., $\Lambda$ is 
$\sim 0.8$ GeV for a dipole regulator, 0.6 GeV for a monopole and 0.4 GeV 
using a $\theta$-function). The sensitivity 
of the extrapolation to the choice 
of the functional form of the FRR is then an additional source 
of systematic error in the final quoted result. 
In the case of the nucleon mass the corresponding 
systematic error was~\cite{Leinweber:2003dg} of the order of a mere 0.1\%.

One of the most remarkable results of this physical understanding of the role 
of the pion cloud and, in particular, 
its suppression at large pion mass has been 
the unexpected discovery of a connection between lattice simulations 
based upon quenched QCD (QQCD) and full QCD~\cite{Aoki:1999yr}. 
In a study of the quark mass dependence of the 
N and $\Delta$ masses~\cite{Young:2002cj},  
it was discovered that if the self-energies appropriate 
to either QQCD or full QCD were regulated using the same 
dipole form for the FRR (the dipole being the most natural physical 
choice given that the axial form factor of the nucleon has a dipole form) with 
mass parameter $\Lambda = 0.8$ GeV 
(the preferred value, as noted above), then the 
residual expansions for the nucleon mass in QQCD and QCD 
(and also for the $\Delta$ 
in QQCD and QCD) were the same within the 
errors of the fit! This is a remarkable 
result which a posteriori gives enormous support to the physical picture of the 
baryons consisting of confined valence quarks surrounded by a perturbative 
pion cloud. The baryon core is basically determined 
by the confinement mechanism 
and provided the choice of lattice scale reproduces the physically known 
confining force (either through the string tension or the 
Sommer parameter~\cite{Sommer:1993ce}, 
derived from the heavy quark potential) 
it makes little difference whether one uses 
QQCD or full QCD to describe that core. 
What {\em does} matter is the change 
in the chiral coefficients as one goes from QQCD to full QCD.

Perhaps the most significant application of 
this discovery has been the application 
to the calculation of the octet magnetic moments and charge radii based on 
accurate QQCD simulations that extend to rather low quark mass. Using 
the constraints of charge symmetry this has led to some extremely accurate 
calculations of the strange quark contributions to 
the magnetic moment~\cite{Leinweber:2004tc} and charge 
radius~\cite{Leinweber:2006ug} of the proton. Indeed, those calculations 
are in excellent agreement with 
the current world data~\cite{Young:2006jc} but, 
in a unique example in modern strong interaction 
physics, they are an order of magnitude more accurate.

\section{Concluding Remarks}
We have briefly reviewed the power of having a simple and intuitive 
understanding of hadron structure in which the pion cloud is a crucial element. 
Time and again the picture has led to new 
discoveries and predictions. It is extremely unlikely that it will cease to 
be either useful or inspiring in the near future~\cite{Strikman:2003gz}, 
as major new facilities, 
such as the 12 GeV Upgrade at Jefferson Lab, enable us to probe hadron and 
nuclear structure in completely new ways.

\section*{Acknowledgements}
I would like to thank the many students and colleagues who have collaborated 
over in the development of the physics described here. 
I am particularly grateful to Wally Melnitchouk for thoughtful 
comments on early drafts of this manuscript. This 
work was supported by the U.~S.~DOE through contract No.~DE-AC-060R23177, 
under which Jefferson Science Associates 
operates Jefferson Lab.


\end{document}